\documentclass[12pt]{article}
\usepackage{esfconf}

\def\beq{\begin{equation}}
\def\eeq{\end{equation}}
\def\be{\begin{equation}}
\def\ee{\end{equation}}
\def\bea{\begin{eqnarray}}
\def\eea{\end{eqnarray}}
\def\p{\partial}
\def\n{\nabla}
\def\d{\displaystyle}

\begin{document}

\title{Interaction of higher spin massive fields with gravity
in string theory}

\authors{V.D.Pershin}

\addresses{
Department of Theoretical Physics, Tomsk State
University, Tomsk 634050, Russia}

\maketitle

\begin{abstract}
Derivations of consistent equations of motion for the massive spin
two field interacting with gravity is reviewed. From the field
theoretical point of view the most general classical action
describing consistent causal propagation  in vacuum Einstein
spacetime is given. It is also shown that the massive spin two field
can be consistently described in arbitrary background by means of
lagrangian equations representing an infinite series in
curvature.  Within string theory equations of motion for the
massive spin two field coupled to gravity is derived from the
requirement of quantum Weyl invariance of the corresponding two
dimensional sigma-model. In the lowest order in string length the
effective equations of motion are demonstrated to coincide with the
general form of consistent equations derived in field theory.
\end{abstract}

\bigskip

In this contribution I review the recent progress~\cite{our} achieved
in description of massive higher spin fields interacting with external
gravity from both pure field theoretical and string theory points of
view.

In general problems of introducing interaction for higher spin fields
is referred to the presence of auxiliary fields in such theories. For
example, free field realizing irreducible representation of the
$D=4$ Poincare group with spin $s$ and mass $m$ is described by a
rank $s$ symmetric traceless transverse tensor
$\phi_{(\mu_1\ldots\mu_s)}$ satisfying the mass shell condition:
\be
(\p^2-m^2) \phi_{\mu_1\ldots\mu_s} =0 {,} \qquad
\p^\mu \phi_{\mu\mu_1\ldots\mu_{s-1}} =0 {,} \qquad
\phi^\mu{}_{\mu\mu_1\ldots\mu_{s-2}} =0 {.}
\label{irrep}
\ee
To reproduce all these equations from a single lagrangian one needs
to introduce auxiliary fields $\chi_{\mu_1\ldots\mu_{s-2}}$,
$\chi_{\mu_1\ldots\mu_{s-3}}$, \ldots, $\chi$ \cite{fierz,singh}.
These symmetric traceless fields vanish on shell but their presence
in the theory provides lagrangian description of the conditions
(\ref{irrep}).  In higher dimensional spacetimes there appear fields
of more complex tensor structure but general situation remains the
same, i.e.  lagrangian description always requires presence of
unphysical auxiliary degrees of freedom.

Arbitrary interaction may create problems of two different kinds.
First of all, in interacting theory auxiliary fields may acquire
dynamics thus violating the correct number of degrees of freedom.
Usually these extra degrees of freedom have kinetic terms with
the wrong sign thus leading to negative energies. In any case
absence of a consistent free limit makes the theory pathological.
Requirement of conservation of the correct number of degrees of
freedom gives severe restrictions on possible types of interaction.

Another problem that may arise in higher spin fields theories is
connected with possible violation of causal properties. This problem
was first noted in the theory of spin 3/2 field in external
fields~\cite{zwanziger} (see also the review~\cite{zwanziger2} and a
recent discussion in~\cite{3/2}). Turning on interaction in theories
of higher spin fields in general changes the characteristic matrix of
equation of motion and there appears possibility of superluminal
propagation.  Such a situation also should be considered as
pathological. The two problems are believed to be connected with each
other in a subtle way, namely, causality is referred to {\em local}
violations of number of degrees of freedom~\cite{3/2}.

Due to these problems no consistent lagrangian describing a
higher spin massive field interacting with itself or with other
dynamical fields is known at the moment. A possible consistent
description of higher spin fields interaction may be given by string
theory which in addition to a finite number of massless excitations
contains  an infinite tower of massive higher spins modes with masses
proportional to the inverse fundamental string length squared. Namely
presence of this infinite tower provides ultraviolet finiteness of
all known perturbatively consistent string models. In the low energy
limit only the lightest string modes are excited and physics is to be
described by some effective theory of a finite number of fields.

The most convenient covariant method of deriving the corresponding
effective field theory is based on non-linear sigma models which
describe propagation of a string in external background fields. The
requirement of quantum conformal invariance in such models is known
to correctly reproduce the equations of motion for massless
background fields~\cite{cfmp}. The inclusion of massive string modes
into this scheme is more difficult because the corresponding
two-dimensional action is non-renormalizable and should include an
infinite number of terms.  Until recently, only linear effective
equations of motion for the massive string modes were
calculated~\cite{massive}.

In the papers~\cite{bflp} it was shown that it is possible to derive
massive fields equations of motion in string theory using ordinary
perturbative quantum calculations. If one is considering the whole
infinite set of massive fields it makes no difference whether the
theory is renormalizable or not. Infinite number of counterterms
needed for cancellation of divergences generating by a specific
massive field in classical action leads to renormalization of an
infinite number of massive fields. The only property of the theory
crucial for possibility of derivation of perturbative information is
that number of massive fields giving contributions to renormalization
of the given field should be finite. To calculate $\beta-$function
for any massive field it is sufficient to find divergences coming
only from a finite number of other massive fields and so it is
possible to derive effective equations of motion for any background
fields in any order.

An apparent disadvantage of such perturbative $\sigma-$model
calculations is that only linear in massive fields equations of
motion can be derived and so one cannot derive cubic interaction for
massive fields. This fact agrees with the old
observation~\cite{versus} that general covariant string field action
describing self-interaction of massive fields cannot exist.

Nevertheless, one can derive quadratic part of effective action for a
single massive field interacting with a collection of massless
fields. Let us now describe how this possibility is explicitly
realized on the example of massive spin 2 field interacting with
external gravity~\cite{our}.

In the flat spacetime the massive spin 2 field is described by
symmetric transversal and traceless tensor of the second rank
$H_{\mu\nu}$ satisfying mass-shell condition:
\be
\Bigr(\partial^2-m^2\Bigl) H_{\mu\nu}=0 {,}\qquad
\partial^\mu H_{\mu\nu}=0 {,}\qquad
H^\mu{}_\mu=0 {.}
\label{irred}
\ee
All these conditions can be derived from the single
lagrangian~\cite{fierz}:
\bea
S&=&\int\! d^D x \biggl\{ \frac{1}{4} \partial_\mu H \partial^\mu H
-\frac{1}{4} \partial_\mu H_{\nu\rho} \partial^\mu H^{\nu\rho}
-\frac{1}{2} \partial^\mu H_{\mu\nu} \partial^\nu H
+\frac{1}{2} \partial_\mu H_{\nu\rho} \partial^\rho H^{\nu\mu}
\nonumber\\&&
\qquad\qquad
{} - \frac{m^2}{4} H_{\mu\nu} H^{\mu\nu} + \frac{m^2}{4}  H^2
 \biggr\}
\label{actfield}
\eea
and $H=\eta^{\mu\nu} H_{\mu\nu}$ plays the role of auxiliary field.

Among the equations of motion
$E_{\mu\nu}=\frac{\d\delta S~~}{\d\delta H^{\mu\nu}}$
there are $D$ primary
constraints $\varphi_\nu^{(1)}=E_{0\mu}$ which do not contain second
time derivatives of the field $H_{\mu\nu}$. Their conservation in
time leads to $D$ secondary constraints
\beq
\varphi_\nu^{(2)} =
\partial^\mu E_{\mu\nu} = m^2 \partial_\nu H - m^2 \partial^\mu
H_{\mu\nu} \approx 0
\label{con1}
\eeq
Conservation of $\varphi_\mu^{(2)}$ defines accelerations $\ddot
H_{0i}$ and leads to one more constraint
\bea
\varphi^{(3)}=
\partial^\mu \partial^\nu E_{\mu\nu}
+ \frac{m^2}{D-2} \eta^{\mu\nu} E_{\mu\nu}
=  H  m^4 \frac{D-1}{D-2} \approx 0
\label{con2}
\eea
Conservation of $\varphi^{(3)}$ gives one more constraint on initial
values and from the conservation of this last constraint the
acceleration $\ddot H_{00}$ is defined. Altogether there are $2D+2$
constraints on the initial values of $\dot H_{\mu\nu}$ and
$H_{\mu\nu}$ and equations of motion are hyperbolic and causal.

Now if we want to construct a theory of massive spin 2 field on a
curved manifold first of all we should provide the same number of
propagating degrees of freedom as in the flat case. It means that new
equations of motion $E_{\mu\nu}$ should lead to exactly $2D+2$
constraints  and in the flat spacetime limit these constraints should
reduce to their flat counterparts.

One possible way to do it is to consider theory in a vacuum Einstein
spacetime:
\be
R_{\mu\nu} = \frac{1}{D} G_{\mu\nu} R
\ee
In this case there exists
one-parameter family of consistent theories with the same number of
constraints as in the flat space:
\bea
&&
S=\int d^D x\sqrt{-G} \biggl\{ \frac{1}{4} \nabla_\mu H \nabla^\mu H
-\frac{1}{4} \nabla_\mu H_{\nu\rho} \nabla^\mu H^{\nu\rho}
-\frac{1}{2} \nabla^\mu H_{\mu\nu} \nabla^\nu H
\\&&
\nonumber
+\frac{1}{2} \nabla_\mu H_{\nu\rho} \nabla^\rho H^{\nu\mu}
+\frac{\xi}{2D} R H_{\mu\nu} H^{\mu\nu} +\frac{1-2\xi}{4D} R H^2
- \frac{m^2}{4} H_{\mu\nu} H^{\mu\nu} + \frac{m^2}{4} H^2
 \biggr\} {.}
\label{curvact}
\eea
Corresponding equations of motion are hyperbolic and causal:
\bea
&&\nabla^2 H_{\mu\nu}
+ 2 R^\alpha{}_\mu{}^\beta{}_\nu H_{\alpha\beta}
+ \frac{2(\xi-1)}{D} R H_{\mu\nu} - m^2 H_{\mu\nu} = 0 {,}
\nonumber\\&&
H^\mu{}_\mu=0 {,} \qquad\qquad\dot H^\mu{}_\mu=0 {,} \qquad\qquad
\nabla^\mu H_{\mu\nu} = 0 {,}
\label{curv_shell}
\eea
provided that parameters of the theory fulfill the conditions:
\be
\frac{2(1-\xi)}{D} R + m^2 \neq 0 {,} \qquad
\frac{D+2\xi(1-D)}{D} R + m^2 (D-1) \neq 0
\label{ineq}
\ee

But can one remove the vacuum Einstein condition and construct
consistent theory for both massive spin 2 field and dynamical
gravity? There are arguments~\cite{duff} against this possibility
based on Kaluza-Klein reduction of $(D+1)-$dimensional gravity. In
string theory there arises another way out of this problem. Namely,
consistent action may be built in a form of infinite series in
curvature or, equivalently, in inverse string length.

From the field theoretical point of view consistent action in the
lowest order in curvature depends on three parameters of
non-minimal coupling:
\bea
&&
S_H =\int d^D x\sqrt{-G} \biggl\{ \frac{1}{4} \nabla_\mu H \nabla^\mu H
-\frac{1}{4} \nabla_\mu H_{\nu\rho} \nabla^\mu H^{\nu\rho}
-\frac{1}{2} \nabla^\mu H_{\mu\nu} \nabla^\nu H
\nonumber
\\&&
{}+\frac{1}{2} \nabla_\mu H_{\nu\rho} \nabla^\rho H^{\nu\mu}
+\frac{\xi_1}{4} R H_{\alpha\beta} H^{\alpha\beta}
-\frac{\xi_1}{4} R H^2
+\frac{1-2\xi_2}{4} R^{\alpha\beta} H_{\alpha\sigma} H_\beta{}^\sigma
\\&&{}
\nonumber
+\frac{\xi_2}{2} R^{\alpha\beta} H_{\alpha\beta} H
+\frac{\xi_3}{2} R^{\mu\alpha\nu\beta} H_{\mu\nu} H_{\alpha\beta}
- \frac{m^2}{4} H_{\mu\nu} H^{\mu\nu} + \frac{m^2}{4} H^2
+ O\Bigl(\frac{1}{m^2}\Bigr)
 \biggr\}
\label{consistent}
\eea
If gravitational field is also subject to some dynamical
equations of the form $R_{\mu\nu}=O(1/m^2)$ then the system
in the lowest order equations of motion look like
\bea
&& \n^2 H_{\mu\nu} - m^2 H_{\mu\nu}
+(\xi_3+2) R_\mu{}^\alpha{}_\nu{}^\beta H_{\alpha\beta}
+ O\Bigl(\frac{1}{m^2}\Bigr) = 0 {,}
\nonumber\\&&
H + O\Bigl(\frac{1}{m^2}\Bigr) = 0 {,} \qquad
\n^\mu H_{\mu\nu} + O\Bigl(\frac{1}{m^2}\Bigr) = 0 {,}
\nonumber\\&&
R_{\mu\nu} + O\Bigl(\frac{1}{m^2}\Bigr) = 0
\label{ricci}
\eea

In string theory equations of motion for such a system can be derived
from the condition of quantum conformal invariance imposed on the
two-dimensional $\sigma-$model with the classical action
\begin{equation}
S=S_0+S_I=
\frac{1}{4\pi\alpha'}\int_M \!\!d^2z\sqrt{g}
      g^{ab}\partial_ax^\mu\partial_bx^\nu G_{\mu\nu}
+\frac{1}{2\pi\alpha'\mu}\int_{\partial M} e dt \;
      H_{\mu\nu}\dot{x}^\mu\dot{x}^\nu
\label{actstring}
\end{equation}
Here $\mu,\nu=0,\ldots,D-1$; $a,b=0,1$,
$\dot{x}^\mu=\frac{dx^\mu}{edt}$. The first term $S_0$ is an integral
over two-dimensional string world sheet $M$ with metric $g_{ab}$ and
the second $S_I$ represents a one-dimensional integral over its
boundary with einbein $e$.

Quantum conformal invariance in the lowest order in $\alpha'$ gives
the conditions
\begin{eqnarray}
&& \nabla^2 H_{\mu\nu}
+ R_\mu{}^\alpha{}_\nu{}^\beta H_{\alpha\beta}
- \frac{1}{\alpha'} H_{\mu\nu} + O(\alpha') = 0 {,}
\nonumber\\&&
\nabla^\alpha H_{\alpha\nu} + O(\alpha') =0 {,} \qquad
H^\mu{}_\mu +O(\alpha') =0 {,}
\nonumber\\&&
R_{\mu\nu} +O(\alpha')= 0 {.}
\label{final}
\end{eqnarray}
which are a particular case of the most general consistent equations
of motion derived from the field theory point of view (\ref{ricci}).

The main implication of this result is that string theory overcomes
the consistency problems of higher spin massive fields interaction by
means of infinite series in $\alpha'$. Correct number of degrees of
freedom can be restored by imposing conditions on the couplings of
the theory in each order. The next interesting step in this direction
would consist in calculation of string loop corrected
$\beta-$functions which due to Fischler-Susskind~\cite{fs} mechanism
should provide full Einstein equations of motion with stress tensor
depending on massive spin 2 field.

\noindent{\bf Acknowledgements.}
I am grateful to I.L. Buchbinder, D.M. Gitman and V.A. Krykhtin for
collaboration in the work reviewed here and to A. Sagnotti,
A.A. Tseytlin and M.A. Vasiliev for stimulating discussions.
The work was supported by grants GRACENAS 97-6.2-34, RFBR
99-02-16617, RFBR-DFG 99-02-04022 and INTAS N 991-590.


\begin{thebibliography}{99}
\bibitem{our}
    I.L. Buchbinder, V.A. Krykhtin, V.D. Pershin,
    {\it Phys. Lett. \bf 466B} (1999) 216;
    I.L. Buchbinder, D.M. Gitman, V.A. Krykhtin, V.D. Pershin,
    {\it Nucl. Phys. \bf B584} (2000) 615;
    I.L. Buchbinder, D.M. Gitman, V.D. Pershin,
    ``Causality of Massive Spin 2 Field in External Gravity'',
    {\it Phys. Lett.} (2000) to appear; {\tt hep-th/0006144}.

\bibitem{fierz}
    M. Fierz, W. Pauli,
    {\it Proc. Royal Soc. \bf A173} (1939) 211.

\bibitem{singh}
   S.J. Chang, {\it Phys.Rev. \bf 161} (1967) 1308;
   L.P.S. Singh, C.R. Hagen, {\it Phys. Rev. \bf D9} (1974) 898.

\bibitem{zwanziger}
   G. Velo, D. Zwanziger, {\it Phys. Rev. \bf 188} (1969) 2218;
   G. Velo, {\it Nucl. Phys. \bf B43} (1972) 389.

\bibitem{zwanziger2}
   D. Zwanziger, {\it Lecture Notes in Physics, \bf 73} (1978) 143.

\bibitem{3/2}
   S. Deser, V. Pascalutsa, A. Waldron,
   ``Massive Spin 3/2 Electrodynamics'', {\tt hep-th/0003011}.

\bibitem{cfmp}
   C. Callan, D. Friedan, E. Martinec, M. Perry,
   {\it Nucl. Phys. \bf B262} (1985) 593;
   E.S. Fradkin,  A.A. Tseytlin,
   {\it Phys. Lett. \bf 158B} (1985) 316;
   {\it Nucl. Phys. \bf B261} (1985) 1.

\bibitem{massive}
   J.M.F. Labastida, M.A.H. Vozmediano,
   {\it Nucl.Phys. \bf B312} (1989) 308;
   I.L. Buchbinder, V.D. Pershin, G.B. Toder,
   {\it Class. Quant. Grav. \bf 14} (1997) 589;
   A. Wilkins,
   {\it Mod. Phys. Lett. \bf A13} (1998) 1289.

\bibitem{bflp}
   I.L. Buchbinder, E.S.  Fradkin, S.L. Lyakhovich, V.D. Pershin,
   {\it Phys.Lett. \bf B304} (1993) 239;
   I.L. Buchbinder, V.A. Krykhtin, V.D. Pershin,
   {\it Phys. Lett. \bf B348} (1995) 63.

\bibitem{versus}
   A.A. Tseytlin, {\it Phys. Lett. \bf B185} (1987) 59.

\bibitem{duff}
   M.J. Duff, C.N. Pope, K.S. Stelle,
   {\it Phys. Lett. \bf B223} (1989) 386.

\bibitem{fs}
   W. Fischler, L. Susskind,
   {\it Phys. Lett. \bf B171} (1986) 383;
   {\it ibid. \bf B173} (1986) 262.

\end{thebibliography}
\end{document}